\documentclass[aip,jap,reprint,twocolumn,amsmath]{revtex4-2}
\usepackage{graphicx}
\usepackage{hyperref}

\begin{document}

\title{Electrical conductivity of nanoring-based transparent conductive films: A mean-field approach}
\author{Yuri Yu. Tarasevich}
\email[Corresponding author: ]{tarasevich@asu.edu.ru}
\author{Andrei V. Eserkepov}
\email{dantealigjery49@gmail.com}
\author{Irina V. Vodolazskaya}
\email{vodolazskaya\_agu@mail.ru}
\affiliation{Laboratory of Mathematical Modeling, Astrakhan State University, Astrakhan 414056, Russia}

\date{\today}

\begin{abstract}
We have studied the electrical conductivity of nanoring-based, transparent conductive films, these being promising elements for flexible electronic devices. Both the wire resistance and the junction resistance were taken into account. We have calculated the dependency of the electrical conductivity on the number density of the rings. We have proposed a mean-field approach to estimate the dependency of the electrical conductivity on the main parameters. Comparison of  direct computations of the electrical conductivity and the estimates provided by the mean-field approach evidenced the applicability of this approach for those cases where the wire resistance dominates over the junction resistance and where both resistances are of the same order. For these two cases, both the direct computations and the mean-field approach evidenced a linear dependence of the electrical conductivity of the films on the number density of the conductive rings. By contrast, the dependence of the electrical conductivity on the number density of the conductive rings is a quadratic when the junction resistance dominates over the wire resistance. In this case, the mean-field approach significantly overestimates the electrical conductivity, since the main assumptions underlying this approach are no longer fulfilled.
\end{abstract}

\maketitle

\section{Introduction}\label{sec:intro}
Transparent electrodes are used in various electronic devices such as touch screens, transparent heaters, displays, and solar cells.\cite{Gao2016AdvPhys,Gupta2016,Haverkort2018,Papanastasiou2020,Zhang2020ChemRev} The widely-used oxide-based devices, e.g., zinc oxide (ZnO), indium tin oxide (ITO), etc., have a number of disadvantages, which have led to an intensive search for alternatives.\cite{Gao2016AdvPhys} Recent advances in nanomaterial research have led to the development of new transparent conductive materials, such as those based on carbon nanotubes (CNTs), graphene, metal nanowires (NWs), and printed metal grids.

The use of nanoring-based conductive films\cite{Layani2009ACSN,Shimoni2014,Azani2018ChEJ} looks extremely attractive, since in this case there are no dead ends in the percolation clusters, i.e., each percolation cluster is identical to its geometric backbone. In the case of transparent heaters, uniform heating is a key requirement.\cite{Gupta2016} Therefore overheated regions (hot spots), which lead to degradation of the conductive films, is a serious problem (see, e.g., Refs.~\onlinecite{Sannicolo2016a,Kumar2017JAP,Khaligh2017,Gupta2017,Sannicolo2018ACSN,Patil2020,Wang2021,Koo2021,Charvin2021}). However, in the case of nanoring-based conductive films, hot spots are less likely.

Nanowire- and nanoring-based conductive films can be treated as random resistor networks (RRNs). The potentials and currents in any RRN can, in principle, be found using Ohm's law and Kirchhoff's rule.\cite{Kirkpatrick1971PRL,Kirkpatrick1973,Yuge1978JPC,Li2007JPhA,He2018JAP,Kim2018JAP,Benda2019JAP,Kim2020JCPC} However, such computations can be very resource-intensive. Another disadvantage of such direct computations of the electrical conductivity is that they give the dependence of the electrical  conductivity on the key physical parameters as graphs or tables rather than as analytical expressions. Various theoretical techniques have been used to estimate the electrical conductivity of RRNs, e.g., percolation theory,\cite{Zezelj2012PRB} effective medium theory,\cite{Callaghan2016PCCP} and the mean-field approach (MFA).\cite{Kumar2016JAP,Kumar2017JAP,Forro2018ACSN}

In the present study, we used direct computations of the electrical conductivity of nanoring-based conductive films along with corresponding MFA estimates. Both the electrical conductivity of rings and the electrical conductivity of the junctions between rings were taken into account. We focued on three limiting cases, viz., the wire-resistance-dominated case (WDR), the junction-resistance-dominated case (JDR), and the case when both the resistances were equal.

The rest of the paper is constructed as follows. Section~\ref{sec:methods} describes some technical details of the simulation and our modification of the mean field approach. Section~\ref{sec:results} presents our main findings. Section~\ref{sec:concl} summarizes the main results. Some of the mathematical details are presented in Appendix~\ref{sec:MFA}.

\section{Methods\label{sec:methods}}
\subsubsection{Sampling\label{subsec:sampling}}
\begin{figure}[!htb]
  \centering
  \includegraphics[width=0.9\columnwidth]{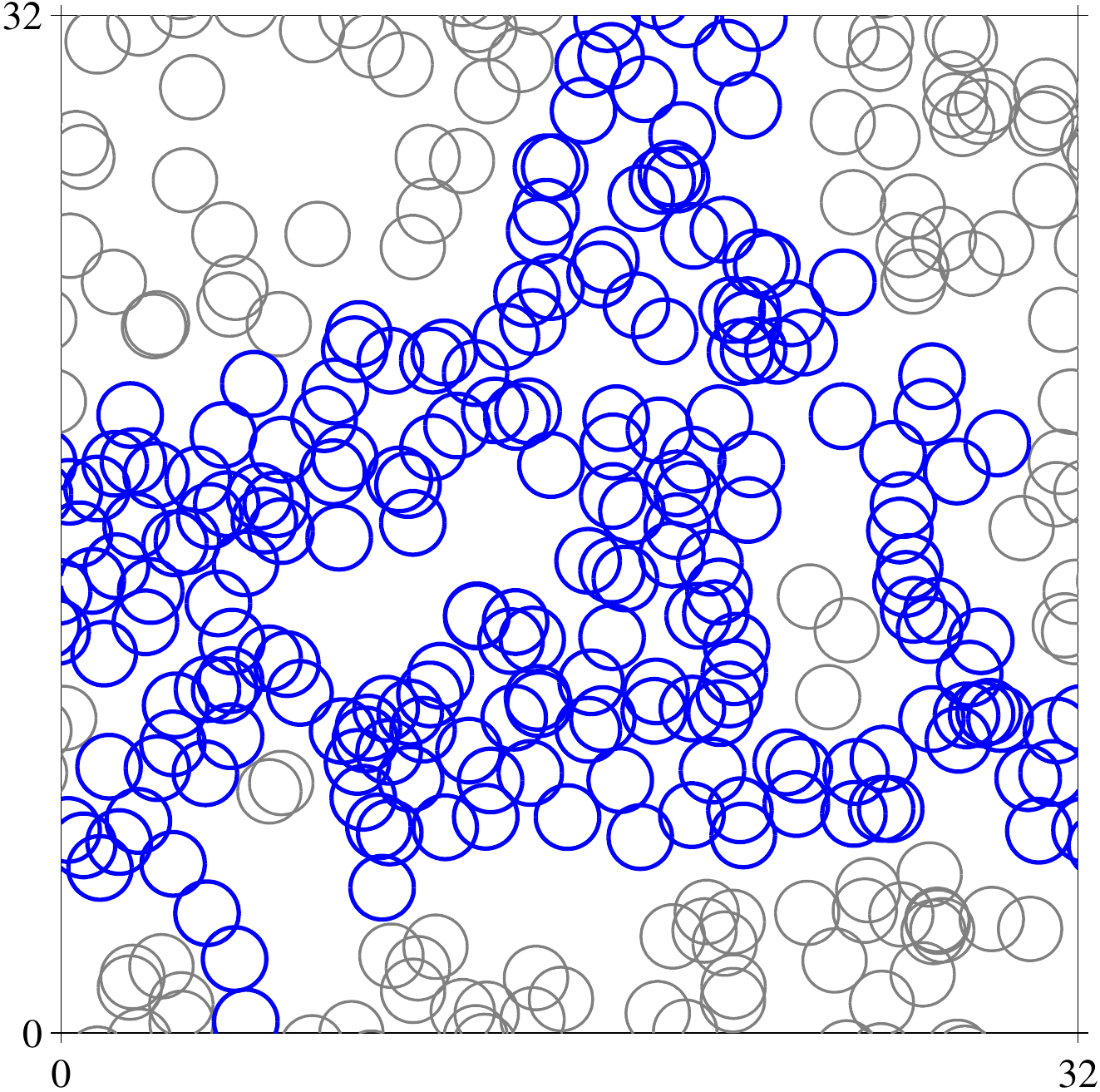}
  \caption{Example of a simulated system exactly at the percolation threshold. The percolation cluster is shown in blue, other clusters are shown in grey.\label{fig:system}}
\end{figure}

We consider $N$ conductive rings of equal radius, $r$, having their centers  independent and identically distributed within a square domain $\mathcal{D}$ of size $L \times L$ with periodic boundary conditions (PBCs). The relation  $r \ll L$ is assumed. The number density of the rings is defined as
\begin{equation}\label{eq:numberdensity}
n = \frac{N}{L^2}.
\end{equation}
We supposed the number density to be larger than the percolation threshold $n > n_\text{c}$. Currently, the best known estimate of the percolation threshold in the thermodynamic limit is $\pi r^2 n_\text{c} \approx 1.12808737$ (see Ref.~\onlinecite{Mertens2012}). When the desired number density of the rings was reached, the PBCs were removed, allowing us to consider the model as an insulating film of size $ L \times L $ covered by conductive rings (Fig.~\ref{fig:system}).

A potential difference, $V_0$, was applied to the opposite borders of this film. The electrical resistance per unit length of  each ring was $R$. The electrical resistance of each contact (junction) between any two rings was $R_\text{j}$.

The probability of intersection of any two rings is
\begin{equation}\label{eq:P}
  P = 4\pi \left( \frac{r}{L}\right)^2.
\end{equation}
Hence, the expected number of junctions is $N_\text{j} = P N^2$, while the expected number of arcs is $N_\text{a} = 2 P N^2$.

We considered three limiting cases, viz., the wire-resistance-dominated (WDR) case when $R=1$, $R_\text{j} = 10^{-6}$, the junction-resistance-dominated (JDR) case when $R=10^{-6}$, $R_\text{j} = 1$, and the case when both the resistances are equal  $R=1$, $R_\text{j} = 1$.  Hereinafter  arbitrary units are used. In our study, $r=1$, $L=32$.

We considered systems with values of the number density up to 5. For each value of the number density, 100 samples were generated. The electrical conductivity was calculated for each sample and then averaged. In the figures, the error bars correspond to the standard deviation of the mean. When not shown explicitly, they are of the order of the marker size.

\subsubsection{Direct computation of the electrical conductivity}\label{subsec:computation}
\begin{figure}[!htb]
  \centering
  \includegraphics[width=0.8\columnwidth]{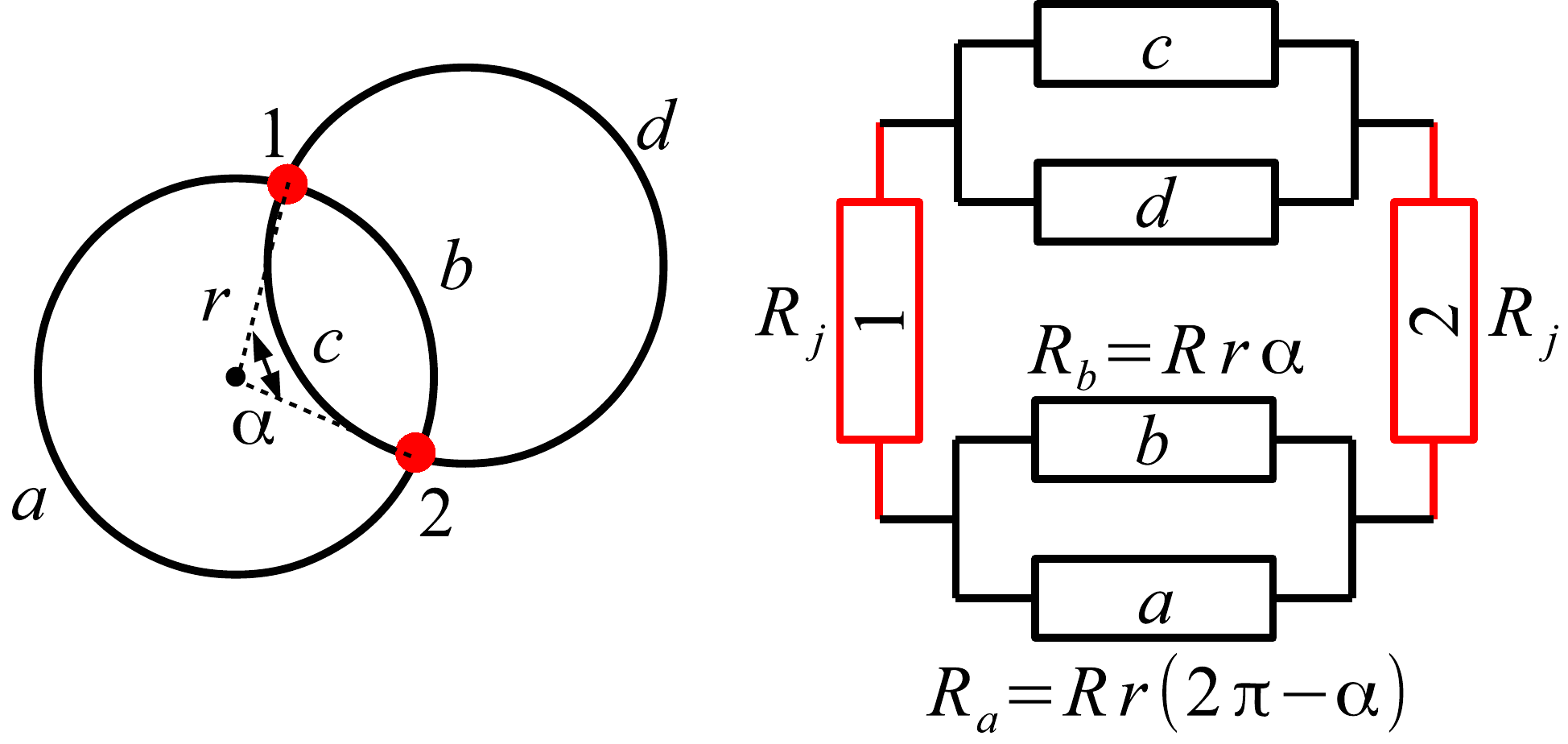}
  \caption{Example of equivalent resistor network corresponding to the two intersecting conductive rings. \label{fig:transformation}}
\end{figure}

The set of intersecting conductive rings can be treated as a random resistor network (RRN). Two kinds of resistors are represented in such a network, viz., resistors corresponding to junctions between two rings, $R_\text{j}$, and resistors corresponding to the ring arcs between two junctions (Fig.~\ref{fig:transformation}).

The total number of resistors under consideration is $N_\text{a} + N_\text{j} = 3 P N^2 = 4 \pi r^2 n^2 L^2$. Applying Ohm's law to each resistor and Kirchhoff's point rule to each node, a system of linear equations (SLE) can be written. The matrix of this SLE is sparse. Such an SLE can be solved numerically. We used the EIGEN library\cite{Guennebaud2010} to solve it. In our study, the SLEs were of orders up to $10^5$ equations.

\subsubsection{Mean-field approach}\label{subsec:MFA}
\begin{figure}[!htb]
  \centering
  \includegraphics[width=\columnwidth]{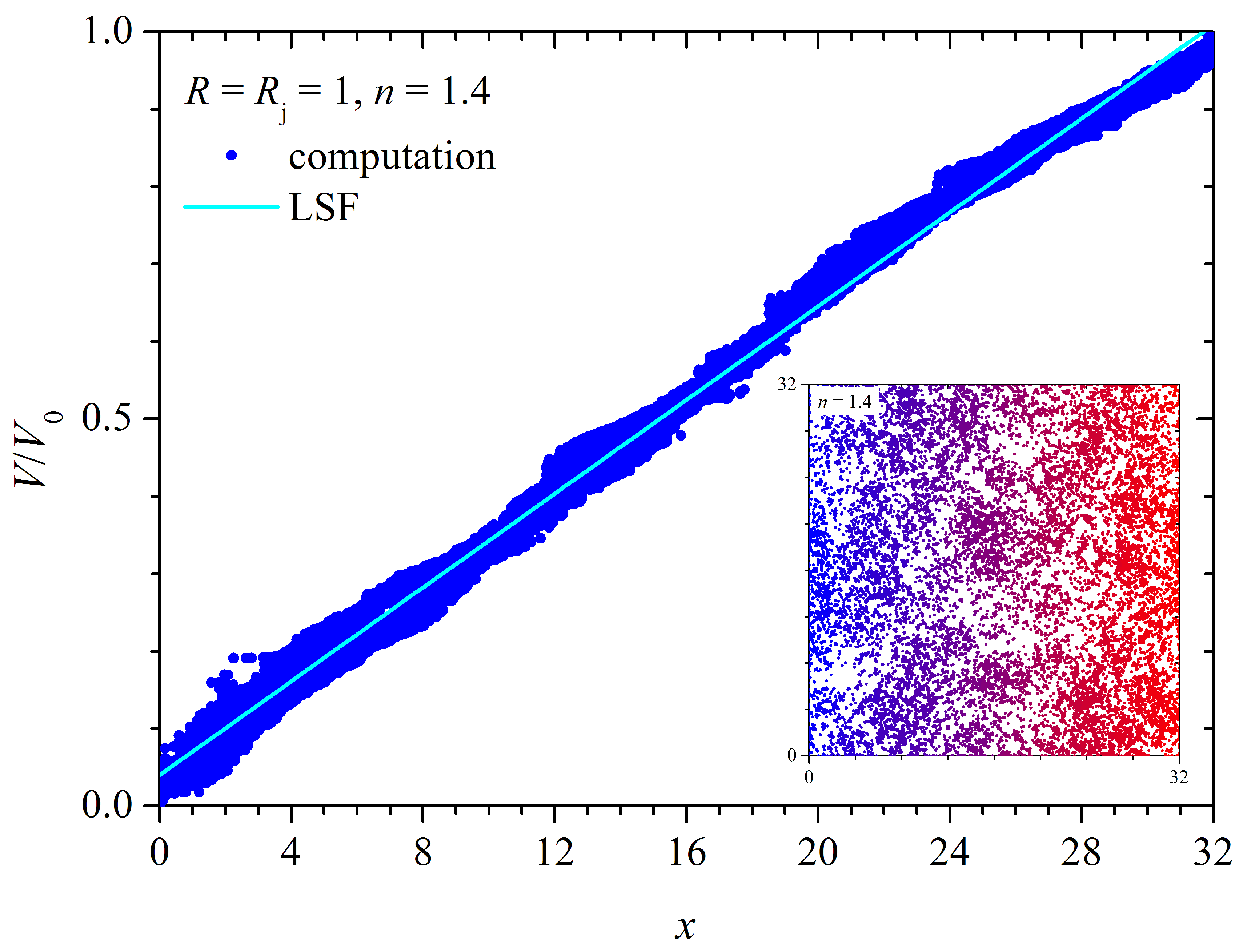}
  \caption{Example of potential distribution in one particular sample at $n=1.4$. The potential of each junction in the system is plotted here against its position in the sample. \label{fig:JWRpotential}}
\end{figure}

When the number density of the conductive rings is large enough, the variation of the electrical potential along the film is close to linear (see, e.g., Ref.~\onlinecite{Azani2019} and Fig.~\ref{fig:JWRpotential}).

Thus, instead of a consideration of the whole RRN, it is possible to study only one ring in the mean-field produced by all the other rings. We present here only the basic formulas. Detailed derivation can be found in Appendix~\ref{sec:MFA}.

\begin{figure}[!htb]
  \centering
  \includegraphics[width=0.5\columnwidth]{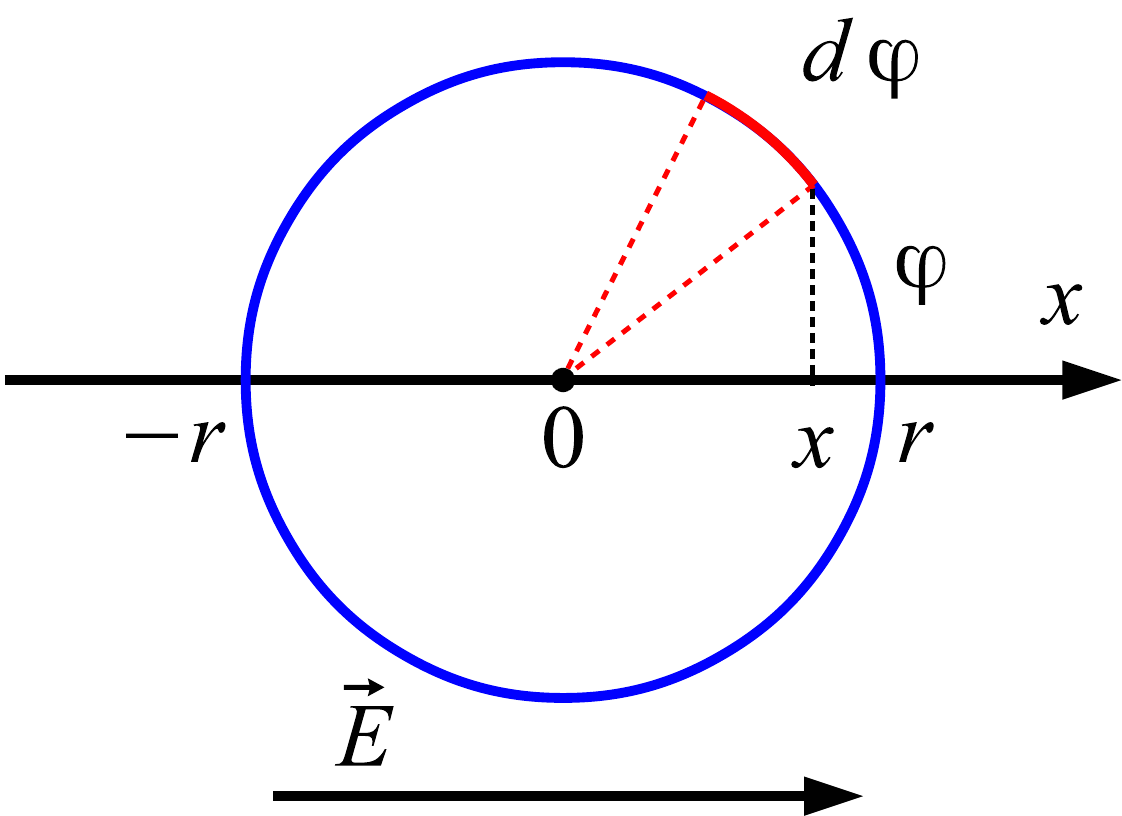}
  \caption{Conductive ring in an external electric field.  In polar co-ordinates, the electric potential is $V(\varphi) = - r E \cos(\varphi)$.\label{fig:ring}}
\end{figure}

Let there be a conductive ring of radius $r$, which is characterized by an electrical resistance $R$ and an insulation leakage conductance $G$, both quantities being referred to the unit length of the ring. The ring is placed in an external field with a coordinate-dependent potential $V(x)$ (Fig.~\ref{fig:ring}).

The electrical current and voltage in this ring are described by a system of ordinary differential equations (ODEs)
\begin{align}
\frac{d u(\varphi)}{d \varphi} + i(\varphi) r R &= 0,\label{eq:dudx}\\
\frac{d i (\varphi)}{d \varphi}  + r G [u(\varphi) - V(\varphi)] & = 0.\label{eq:didx}
\end{align}
Here, $V(\varphi) = - r E \cos \varphi $.
These ODEs can be treated as a kind of telegraph equations for direct current.
The solution of the system of ODEs~\eqref{eq:dudx} and \eqref{eq:didx} is
\begin{equation}\label{eq:i2}
i(\varphi) = -\frac{\lambda^2 r^2 E\sin \varphi}{R(\lambda^2 r^2 +1)},
\end{equation}
\begin{equation}\label{eq:u2}
u(\varphi) =  - \frac{\lambda^2 r^3 E\cos \varphi}{\lambda^2 r^2 +1},
\end{equation}
where $\lambda^2 = RG$.
The dependence of the current on coordinate $x$ is
\begin{equation}\label{eq:ixsolution}
i(x) = \frac{\lambda^2 r^2 E}{R(\lambda^2 r^2 +1)}\sqrt{1-\left(\frac{x}{r}\right)^2}.
\end{equation}

Now, we can turn to a system of $N$ conductive rings. The total electrical current, which is carried by all the rings intersecting a given equipotential line, is
\begin{equation}\label{eq:totalcurrent}
\mathcal{I} =  \frac{\pi n\lambda^2 r^3}{R(\lambda^2 r^2 +1)} V_0.
\end{equation}
Hence, the electrical conductivity of the film is
\begin{equation}\label{eq:sigma-lambda}
\sigma =\frac{\pi \lambda^2 r^3 n}{R(\lambda^2 r^2 +1)}.
\end{equation}

We treat junctions between the rings as representing the leakage conductance. The probability that a given ring intersects exactly $k$ other rings is described by the binomial distribution $p(k) = \binom{k}{m} P^k (1- P)^{N-1-k}$. The expected number of intersections is $\langle k \rangle = 4 \pi n r^2$, since $N \gg 1$. Hence, each ring has, on average, $8 \pi n r^2$ contacts. When the resistance of an individual contact is $ R_\text{j}$, the leakage conductance per unit length is
\begin{equation}\label{eq:leakage}
G =  \frac{4 n r}{R_\text{j}}.
\end{equation}
Hence,
\begin{equation}\label{eq:lambda}
\lambda^2 = \frac{4 n r R}{R_\text{j}}.
\end{equation}
Thus,
\begin{equation}\label{eq:sigma}
\sigma = \frac{4 \pi r^4 n^2}{4 n r^3 R + R_\text{j}}.
\end{equation}

\section{Results}\label{sec:results}
\begin{figure}[!htb]
\centering
\includegraphics[width=\columnwidth]{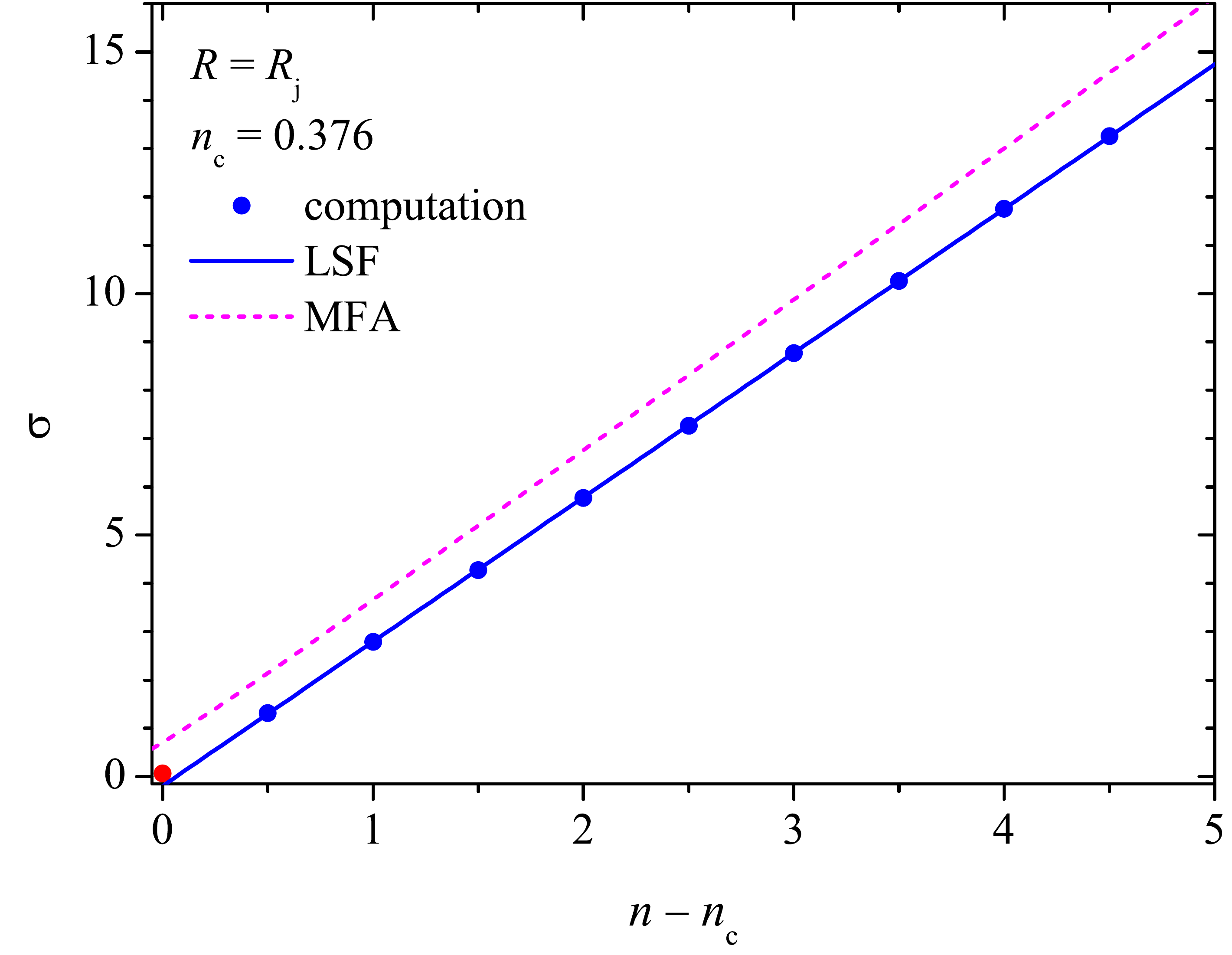}\\
\caption{Dependency of the electrical conductivity on the number density of the conductive
rings when the wire resistance and the junction resistances are equal. The results of the direct computations are shown as closed circles. The least squares fitting (LSF) is shown as a solid line. The MFA prediction is shown as a short-dashed curve.\label{fig:JWR}}
\end{figure}

Figure~\ref{fig:JWR} presents the linear dependency of the electrical conductivity on the number density of the conductive rings when the wire resistance and the junction resistances are equal. The least squares fitting (LSF) with a slope $2.987 \pm 0.003$ is shown as a solid line. The nearest point to the percolation threshold has been excluded from the fitting. The adjusted coefficient of determination $R^2=0.99999$. The MFA prediction is shown as the short-dashed curve, which is very close to a line with a slope~$\pi$. Here, the MFA slightly overestimates the electrical conductivity.

\begin{figure}[!htb]
\centering
\includegraphics[width=\columnwidth]{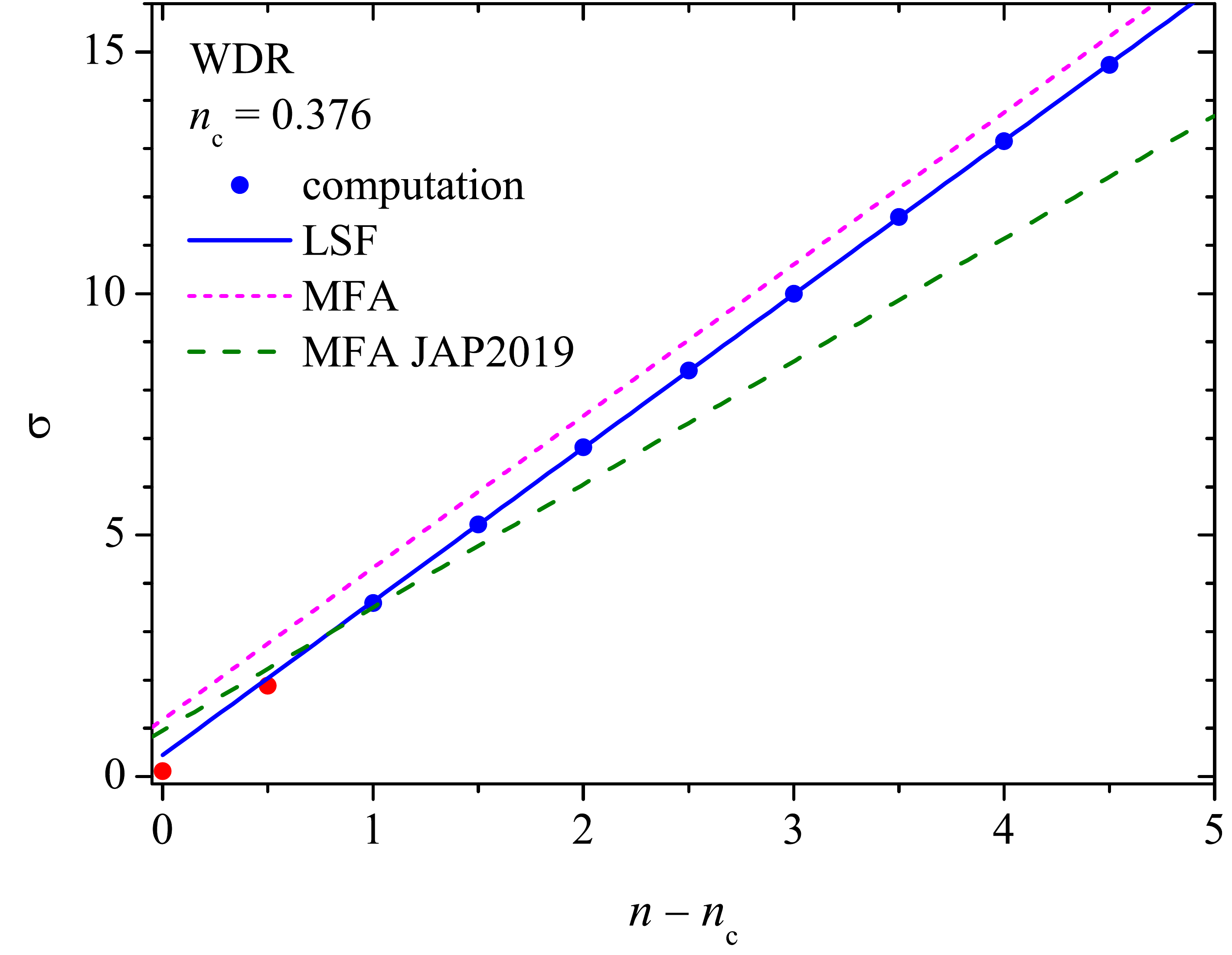}\\
\caption{Dependency of the electrical conductivity on the number density of the conductive
rings for the wire-resistance-dominated case. The results of the direct computations are shown as closed circles. The least squares fitting (LSF) is shown as a solid line. Our MFA prediction is shown as the short-dashed curve. The prediction of alternative kind of MFA\cite{Azani2019} is shown as the dashed line.\label{fig:WDR}}
\end{figure}

Figure~\ref{fig:WDR} demonstrates the linear dependency of the electrical conductivity on the number density of the conductive rings for the wire-resistance-dominated case. The least squares fitting (LSF) with a slope $3.179 \pm 0.006$ is shown as a solid line. Two nearest points to the percolation threshold have been excluded from the fitting. The adjusted  coefficient of determination $R^2=0.99998$. The MFA prediction is shown as the short-dashed curve, which is very close to a line with a slope~$\pi$. Again, the MFA slightly overestimates the electrical conductivity.  For comparison, the prediction using a different kind of MFA\cite{Azani2019} is shown as the dashed line. The latter kind of MFA is based on a geometrical consideration.\cite{Kumar2016JAP} This approach has been successfully applied to different kinds of conductive films,\cite{Kumar2017JAP,Gupta2017,Kumar2019IEEE} however, for the particular film under consideration, its prediction is less accurate than that provided by our variant of the MFA.

\begin{figure}[!htb]
\centering
\includegraphics[width=\columnwidth]{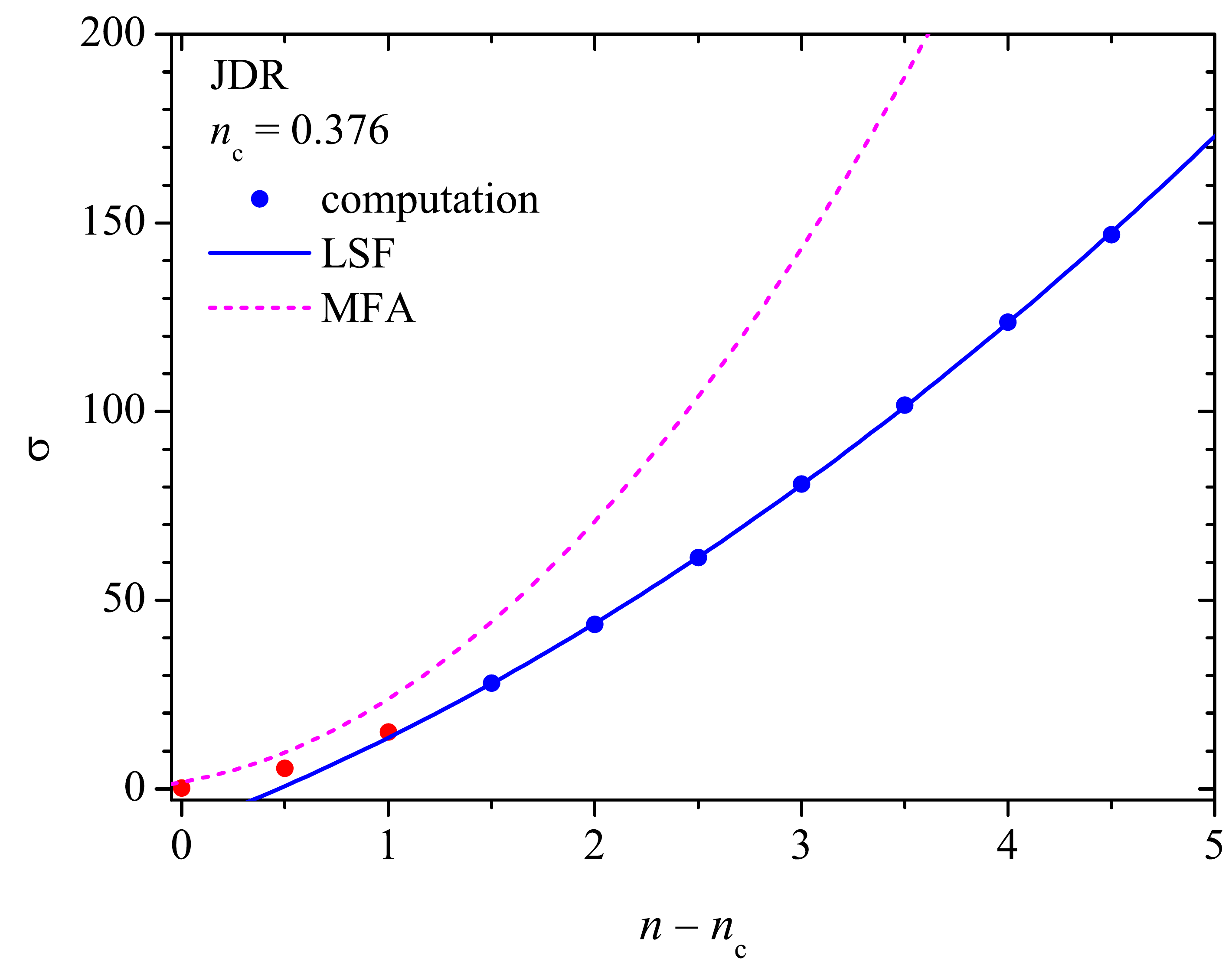}\\
\caption{Dependency of the electrical conductivity on the number density of conductive the
rings for the junction-resistance-dominated case. The results of the direct computations are shown as closed circles. The least squares fitting (LSF) is shown as the solid curve. Our MFA prediction is shown as the short-dashed curve.\label{fig:JDR}}
\end{figure}

Figure~\ref{fig:JDR} shows the dependency of the electrical conductivity on the number density of the conductive rings for the junction-resistance-dominated case. The least squares fitting (LSF) is shown as the solid curve. A second-order polynomial was used to fit the dependency. The coefficient at the highest degree of the polynomial is $3.17 \pm 0.17$. The three nearest points to the percolation threshold were excluded from the fitting. The adjusted coefficient of determination $R^2=0.99991$. The MFA prediction is shown as the short-dashed curve. Here, the MFA significantly overestimates the electrical conductivity.

\begin{figure}[!htb]
\centering
\includegraphics[width=\columnwidth]{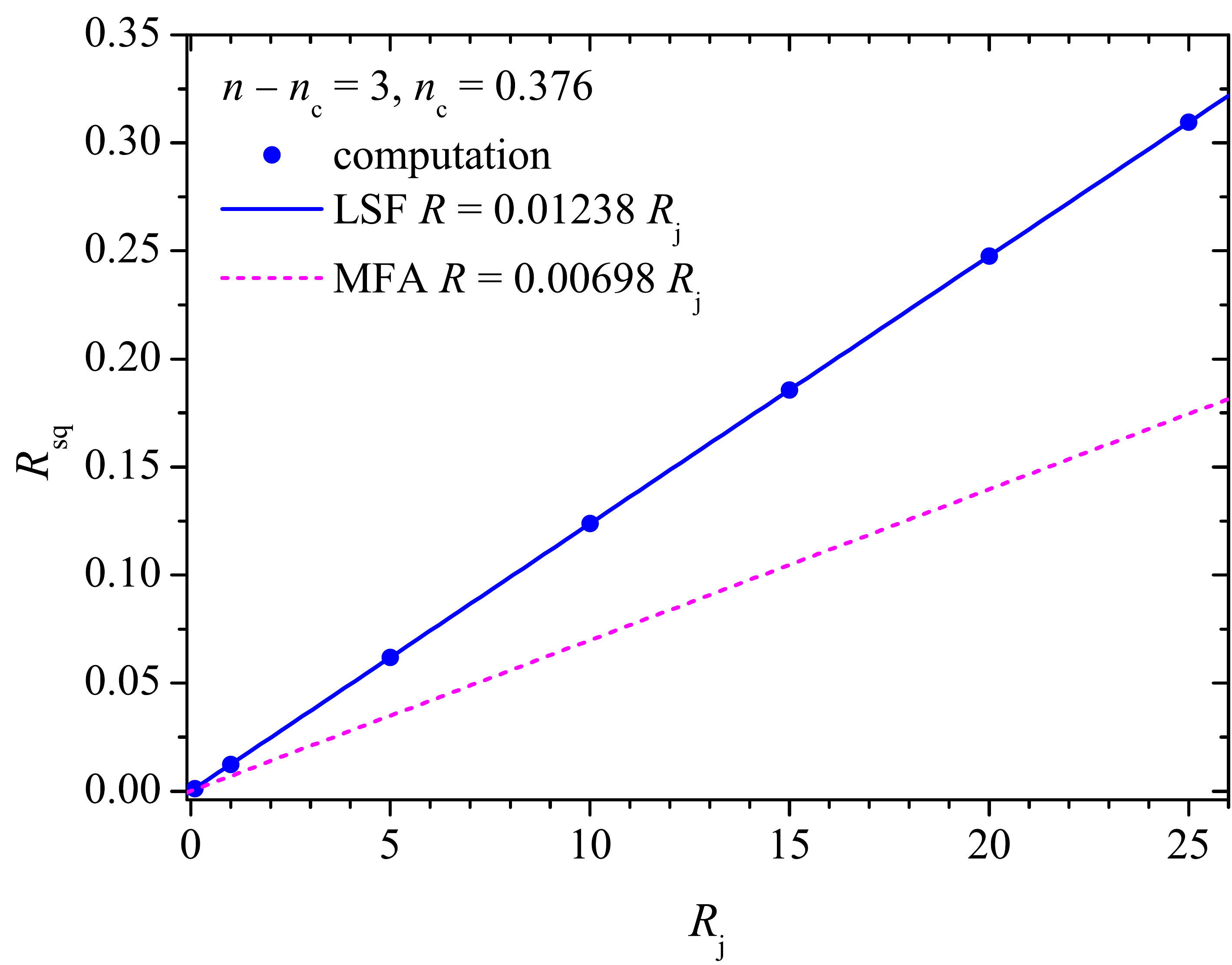}\\
\caption{Dependency of the electrical conductivity on the junction resistance for the junction-resistance-dominated case ($R = 10^{-6}$), when the number density of the conductive rings is fixed ($n - n_\text{c} =3$). The results of the direct computations are shown as closed circles. The least squares fitting (LSF) is shown as the solid line. Our MFA prediction is shown as the  short-dashed line.\label{fig:RvsRj}}
\end{figure}

Figure~\ref{fig:RvsRj} exhibits the dependency of the electrical resistance on the junction resistance for the junction-resistance-dominated case,  when the number density of the conductive rings is fixed. The least squares fitting (LSF) with a slope $0.01238$ is shown as the solid line. The adjusted coefficient of determination $R^2=1$. The MFA prediction is shown as the short-dashed line with a slope $(4\pi)^{-1} \approx 0.00698$. It appears that the slopes differ by  $\approx \sqrt{\pi}$ times. This discrepancy is not surprising. The essential assumptions of the MFA are (i) the system under consideration is planar and (ii) the conductive rings provide a significant contribution to the resistance. Both assumptions were directly used to estimate the total current. Let us imagine the limiting junction-resistance-dominated case when the rings are superconductive ($R = 0$), while the junctions have a finite resistance. In this case, the rings do not contribute to the electrical resistance. By contrast, the contacts completely determine the resistance of the system.   Obviously, an equivalent RRN that contains only resistors corresponding to junctions, is not planar. Thus, neither of the significant assumptions is met in this case.

\section{Conclusion\label{sec:concl}}
We calculated the electrical conductance of two-dimensional systems of randomly placed equally-sized conductive rings accounting for both the resistance of the rings themselves and the resistance of the junctions between the rings. The number density of the rings varied from the percolation threshold up to 5. We found a linear dependence of the conductance on the number density for the wire-resistance-dominated case as well for the case when both the ring resistance and the junction resistances were of the same order. By contrast, a quadratic dependency was observed for the junction-resistance-dominated case.

We applied a mean-field approach to estimate the electrical conductivity. Within this approach, the sheet resistance is
$$
R_\text{sq} = \frac{R}{\pi n r} + \frac{R_\text{j}}{4\pi r^4 n^2},
$$
since we considered only square samples where  $R_\text{sq} = \sigma^{-1}$.
For the wire-resistance-dominated case as well for the case when both the ring resistance and the junction resistances are of the same order, the predictions of the mean-field approach are fairly close to the results of the direct calculations. However, the mean-field approach slightly overestimate the electrical conductivity. Nevertheless, for the  wire-resistance-dominated case, the newly proposed approach offers a better estimate of the electrical conductivity when compared to the approach used in Ref.~\onlinecite{Azani2019}. To our surprise, the mean-field approach correctly predicted the behavior of the electrical conductivity, qualitatively, even in the junction-resistance dominated case, although, in this case, the basic assumptions that were used to derive the formula are not satisfied.

The observed behavior is consistent with the experimental data for random nanowire-networks. Thus, based on an analysis of such experimental data, a simple relation has recently been proposed
$$
R_\text{sq} \approx \alpha\left( R_n + R_j\right),
$$
where $ R_n $ is the average value of the electrical resistance between two contacts, while $\alpha$ is an adjustable parameter.\cite{Ponzoni2019APL}

Moreover, the observed behavior is consistent with predictions obtained using a quite different approach. Thus, the sheet resistance of dense networks of randomly spaced rods can be calculated as
$$
 R = \frac{b n^{t-1} R_s + (n + n_c)^{t-2}R_j}{a\left[(n-n_c)^t + c(L/l)^{-t/\nu}\right]},
$$
where $R_s$ is the resistance of the rod, $a$, $b$, and $c$ are adjustable parameters, and $t$ is the conductivity exponent.\cite{Zezelj2012PRB} However, there is not any adjustable parameter in our formula~\eqref{eq:sigma}.

\begin{acknowledgments}
Y.Y.T. and A.V.E. would like to acknowledge the funding received from the Foundation for the Advancement of Theoretical Physics and Mathematics ``BASIS'', grant~20-1-1-8-1.
\end{acknowledgments}

\appendix
\section{Derivation of the electrical conductivity}\label{sec:MFA}
Consider an arc of the conductive ring that is located between two points with angular co-ordinates $\varphi$ and $\varphi + d\varphi$. In this arc, according to Ohm's law,
\begin{equation*}
d  u(\varphi) + i(\varphi) r R d \varphi = 0,
\end{equation*}
since there is no electromotive force.
In this arc, any change in the electrical current is caused by a loss of charge due to imperfect insulation (leakage conductance of the isolator)
$$
d i (\varphi) + [u(\varphi) - V(\varphi)] r G d\varphi = 0.
$$
Dividing both equations by $d \varphi$, we obtain the system~\eqref{eq:dudx} and~\eqref{eq:didx}.

Differentiating equation~\eqref{eq:dudx} with respect to $\varphi$ and substituting equation~\eqref{eq:didx} into it, we obtain an inhomogeneous linear ODE of the second order
\begin{equation}\label{eq:inhomod2udx2}
\frac{d^2 u(\varphi)}{d \varphi^2} - r^2 R G u(\varphi) = -r^2 R G V(\varphi).
\end{equation}

A solution of the inhomogeneous ODE of the second order~\eqref{eq:inhomod2udx2} equates to the sum of the general solution, $u_0(\varphi)$, of the homogeneous ODE $u''(\varphi) - r^2 R G u(\varphi) = 0$ and a particular solution,  $u^\ast(\varphi)$, of the inhomogeneous ODE.
$$
u_0(\varphi) = A_1 \exp(\lambda r \varphi) + A_2 \exp(-\lambda r \varphi).
$$
Since $V(\varphi) = - r E \cos \varphi $, $u^\ast(\varphi) = A_3 \cos \varphi$. Since the solution have to be periodic, $A_1=A_2=0$. Thus, we obtain~\eqref{eq:u2}.

Let the center of a ring be located at a distance $x$ from an equipotential line ($x<r$). The total current at the two points of intersection of the ring with the equipotential line is
$$
i(x) = 2\frac{\lambda^2 r^2 E}{R(\lambda^2 r^2 +1)}\sqrt{1-\left(\frac{x}{r}\right)^2}.
$$
The number of rings whose centers are located at a distance of $ [x; x + dx] $ from the equipotential line is
$n L dx,$
while the total current that they carry through the equipotential line equals
$$
di = 2\frac{\lambda^2 r^2 E}{R(\lambda^2 r^2 +1)}\sqrt{1-\left(\frac{x}{r}\right)^2}n L dx.
$$
Integration over $x$ leads to formula~\eqref{eq:totalcurrent}.
\bibliography{rings}

\end{document}